# Stoichiometry of Electrostatic Complexes Determined by Light Scattering

**J.-F. Berret**[@]
Matière et Systèmes Complexes, UMR 7057 CNRS Université Denis Diderot Paris-VII, 140 rue de Lourmel, F-75015 Paris France

**Abstract :** We report on the electrostatic complexation between oppositely charged polymers and inorganic nanoparticles investigated by static and dynamical light scattering. The nanoparticles put under scrutiny were citrate-coated nanocrystals of cerium oxide ($CeO_2$, nanoceria), of iron oxide ($\gamma$-$Fe_2O_3$, maghemite) and of europium-doped yttrium vanadate (Eu:$YVO_4$) with sizes in the 10 nm range. For the polymers, we have used cationic-neutral diblock copolymers (poly(trimethylammonium ethylacrylate)-*b*-poly(acrylamide), hereafter referred to as PTEA-*b*-PAM) with different molecular weights. For the three colloidal dispersions, we show that the electrostatic complexation gives rise to the formation of stable nanoparticle clusters in the 100 nm range. The complexation was monitored by systematic measurements of the scattering intensity *versus* X, the mixing ratio between nanoparticles and polymers. For 5 nanoparticle/polymer pairs, namely $CeO_2$/PTEA$_{5K}$-*b*-PAM$_{30K}$, $\gamma$-$Fe_2O_3$/PTEA$_{5K}$-*b*-PAM$_{30K}$, $\gamma$-$Fe_2O_3$/PTEA$_{11K}$-*b*-PAM$_{30K}$, Eu:$YVO_4$/PTEA$_{2K}$-*b*-PAM$_{60K}$ and Eu:$YVO_4$/PTEA$_{5K}$-*b*-PAM$_{30K}$, we found a unique behavior : the scattering intensity exhibits a sharp and prominent peak in the intermediate X-range. To account for this behavior, we have developed a model which assumes that regardless of X, the mixed aggregates are formed at a fixed polymer-to-nanoparticle ratio. The agreement between the results and the model is excellent on the 5 systems. Results at different molecular weights suggest that the stoichiometry of the mixed aggregates is controlled by the electrostatic interactions between the opposite charges. The model allows to derive the molecular weight and the stoichiometry of the mixed aggregates.

# I - Introduction

The electrostatic complexation between oppositely charged macromolecules and colloids has attracted much attention during the last years because it is at the origin of many fundamental non-specific association mechanisms relevant for biological systems. One of the most well-known examples is that of chromatin which is a dense complex of DNA and histone proteins and which is the lowest level of hierarchy of DNA folding of chromosomes [1,2]. Electrostatic complexation is also utilized in applications, e.g. in formulation of personal care products, in coating and composite technologies as well as in water treatment and filtration.

Concerning the complexation mechanism, it has been shown that the driving forces for association are both enthalpic and entropic in origin [3-7]. The enthalpic part in the free energy of association is linked to the pairing of the opposite charges. As demonstrated recently by Laugel and coworkers [7], the binding enthalpy depends strongly on the chemical constituents that are paired in the complexation process. The entropic contribution to the free energy arises from the release of the counterions which are condensed on the surface of the colloid or along the backbone of the polymer [8], as well as from the loss of translational and rotational degrees of freedom of macromolecules in their bound state [9]. The balance between the enthalpic and entropic contributions give rise to wide variety of complexation behaviors. The association can be strong, and then yields the formation of a coacervate and a macroscopic phase separation [10-13]. In some systems, the complexation is weaker and the electrostatic complexes remain soluble [6,14-17].

Since the pioneering work by Bungenberg de Jong on gelatin and arabic gum in the 1940's [18], the complexation has been investigated on various systems, comprising synthetic [19] and biological [16,20-22] polymers, multivalent counterions [23,24], surfactant micelles [13,25-27], organic and inorganic nanoparticles [28-35]. For electrostatic complexes either in the separated or soluble state, it is essential to know their microstructures and stoichiometry, since these properties will ultimately determine the range of applications of these systems. In the present paper, we have studied the complexation behavior between inorganic nanoparticles and oppositely charged polymers. The nanoparticles and polymers investigated here were both considered as strongly charged *i.e.* the distance between neighboring structural charges was of the order or smaller than the Bjerrum length $\lambda_B$ (0.7 nm in water) [36]. The electrostatic interactions between such highly charged systems yielded a macroscopic phase separation. In order to avoid the phase separation, we have used charged-neutral block copolymers instead of homopolyelectrolytes. Doing so, the electrostatic complexation resulted in the formation of core-shell colloids in the 100 nm range. The cores of the colloids were found to be microseparated phases containing tens to hundreds of particles glued together by the polyelectrolyte blocks. Our objective here was to shed some light of the complexation mechanisms through the study of the microstructure of the associated colloids.

In order to demonstrate the generality of this approach, the complexation has been studied using three different types of colloidal dispersions. These particles are nanocrystals of cerium oxide ($CeO_2$, nanoceria), of iron oxide ($\gamma$-$Fe_2O_3$, maghemite) and of europium-doped ytttrium vanadate (Eu:$YVO_4$). These particles were considered because of their potential applications in coating and display technologies [37,38], as well as in biology [39,40]. The complexation was followed by elastic and quasielastic light scattering as a function of the mixing ratio X. For 5 nanoparticle/polymer pairs investigated, we have found a unique behavior : the scattering intensity exhibits a sharp and prominent peak in the intermediate X-range. We have developed a model which assumes that the mixed aggregates are formed at a fixed polymer-to-nanoparticle ratio, regardless of X. The agreement between the results and the model is excellent on the 5 systems. The model allows to derive the molecular weight and the stoichiometry of the mixed aggregates.

## II - Experimental

The nanocrystals of cerium oxide and europium-doped yttrium vanadate were produced by precipitation of rare-earth complexes under controlled thermodynamic conditions [34,37,38]. The colloidal dispersions were kindly provided to us by Rhodia. The iron oxide nanoparticles were obtained by alkaline co-precipitation of iron II and iron III salts and were sorted according to size by successive phase separations [41,42]. The iron oxide batches were made available to us by the Laboratoire des Liquides Ioniques et Interfaces Chargées, Université Pierre et Marie Curie (Paris, France). For the present study, the size and morphology of the particles were characterized by transmission electron microscopy and light scattering. Electron microscopy have shown that the nanoceria consist of isotropic agglomerates of 2 - 5 crystallites with typical size 2 nm. With the technique of cryo-TEM, Eu:YVO4 particles have appeared as anisotropic agglomerates made from the assembly of 8 nm crystallites. The iron oxide particles on the other hand exhibited a more homogenous and spherical microstructure and their diameters were described by log-normal distribution with a most probable value of 6.3 nm and a polydispersity of 0.23 [43]. With static and dynamic light scattering, the weight-average molecular weight $M_w^{Part}$ and the hydrodynamic diameter $D_H$ of the particles were determined. The following values were obtained for cerium oxide, iron oxide and of europium-doped ytttrium vanadate, respectively: $M_w^{Part}$ = 1.47×10$^5$, 3.40×10$^5$ and 2.45×10$^6$ g·mol$^{-1}$, and $D_H$ = 10, 11 and 35 nm (Table I).

| | specimens | $M_w$ Kg mol$^{-1}$ | dn/dc cm$^3$ g$^{-1}$ | K cm$^2$ g$^{-2}$ | $D_H$ nm |
|---|---|---|---|---|---|
| *polymers* | PTEA$_{2K}$-b-PAM$_{60K}$ | 62 000 | | | 19 |
| | PTEA$_{5K}$-b-PAM$_{30K}$ | 35 000 | 0.15 | 0.46×10$^{-6}$ | 13 |
| | PTEA$_{11K}$-b-PAM$_{30K}$ | 41 000 | | | 11 |
| *nanoparticles* | CeO$_2$ | 147 000 | 0.24 | 1.20×10$^{-6}$ | 10 |
| | γ-Fe$_2$O$_3$ | 340 000 | 0.18 | 0.66×10$^{-6}$ | 11 |
| | Eu:YVO$_4$ | 2 450 000 | 0.12 | 0.30×10$^{-6}$ | 35 |

**Table I**
Molecular weight ($M_w$), refractive index increment (dn/dc), light scattering coupling constant K (at 488 nm) and hydrodynamic diameter ($D_H$) characterizing the cationic polymers and the citrate-coated nanoparticles investigated in the present work.

At the pH values at which the complexation occurred (pH 7 – 8), the particles were stabilized by electrostatic interactions mediated by charged ligands. For the three systems, the ligands adsorbed on the surface of the particles were citric acid in its sodium salted form (sodium citrate). The coating of the water-solid interfaces by citrate molecules represents an important step of the synthesis process, since it ensures the stability of the sols in the pH range required for applications [44]. For the three systems, ζ-potential measurements (Zetasizer Nano ZS, Malvern Instrument) were performed and have shown that coated nanoparticles were negatively charged, and thus of opposite charge with that of the polyelectrolyte block [43]. Note that small adsorbing molecules are characterized by adsorption isotherms, i.e. the adsorbed species are in equilibrium with free molecules dispersed in the bulk solution. In the present work, the concentration of free ligands was kept to its minimum in order to avoid residual complexation with the cationic-neutral copolymers [45]. The three anionic citrate-coated particles have been complexed with a cationic-neutral diblock copolymer, referred to as poly(trimethylammonium ethylacrylate)-*b*-poly(acrylamide). The counterion associated with the quaternary ammonium group is methyl sulfate [46]. The diblock copolymers were synthesized by controlled radical polymerization according to MADIX technology [47] and the chemical formulae of the monomers are given in Refs. [48,49]. Three molecular weights were put under scrutiny, corresponding to 7 (2 000 g·mol$^{-1}$), 19 (5 000 g·mol$^{-1}$) and 41 (11 000 g·mol$^{-1}$) monomers in the charged blocks and 420 or 840 (30 000 g·mol$^{-1}$ or 60 000 g·mol$^{-1}$) for the neutral chain. In the following, the copolymers are abbreviated as PTEA$_{2K}$-*b*-PAM$_{60K}$, PTEA$_{5K}$-*b*-PAM$_{30K}$ and PTEA$_{11K}$-*b*-PAM$_{30K}$. The role of the neutral chains was to prevent the coacervate microphase to undergo a precipitation.

Polymer-nanoparticle complexes were obtained by mixing stock solutions prepared at the same weight concentration (c = 0.2 wt.%) and same pH (pH 8). The mixing of the two initial solutions was characterized by the ratio $X = V_{Part}/V_{Pol}$, where $V_{Part}$ and $V_{Pol}$ are the volumes of the particle and polymer solutions respectively. This procedure [46,48] was preferred to titration experiments because it allowed to explore a broad range in mixing ratios ($X = 10^{-2} - 100$) and simultaneously to keep the total concentration in the dilute regime [50]. Doing so, the colloidal interactions between the different species (polymers, nanoparticles or polymer-nanoparticle aggregates) could be neglected. As far as the kinetics is concerned, the formation of the aggregates occurred very rapidly on mixing, i.e. within a time scale inferior to one second. Moreover, in the range of concentration investigated, the mixing procedures were fully reproducible.

*Static and dynamic light scattering*
Static and dynamic light scattering were performed on a Brookhaven spectrometer (BI-9000AT autocorrelator, λ = 488 nm) and on a on a Malvern-Amtec Macrotron spectrometer (λ = 633 nm) for measurements of the Rayleigh ratio $\mathcal{R}(q,c)$ and of the collective diffusion constant D(**c**). The Rayleigh ratio was obtained from the scattered intensity I(q,c) measured at the wave-vector q according to :

$$\mathcal{R}(q,c) = \mathcal{R}_{std} \frac{I(q,c) - I_S}{I_{Tol}} \left(\frac{n}{n_{Tol}}\right)^2 \qquad (1)$$

In Eq. 1, $\mathcal{R}_{std}$ and $n_{Tol}$ are the standard Rayleigh ratio and refractive index of toluene, $I_S$ and $I_{Tol}$ the intensities measured for the solvent and for the toluene in the same scattering configuration and $q = \frac{4\pi n}{\lambda}\sin(\theta/2)$ (with n the refractive index of the solution and θ the scattering angle). Because of the absorption of the incident light due to the rust-colored iron oxide sols, the transmittance at 633 nm was measured by UV-visible spectrometry for the γ-$Fe_2O_3$ solutions and the scattered intensities were corrected accordingly. Absorbance corrections were not necessary for $CeO_2$ and Eu:$YVO_4$ nanosols. Light scattering was used to determine the apparent molecular weight $M_{w,app}$ and radius of the gyration $R_G$ of the macromolecules and colloids investigated here. In the regime of weak colloidal interactions, the Rayleigh ratio $\mathcal{R}(q,c)$ was found to follow a wave-vector and concentration dependence which is highlighted by the Zimm representation [51]:

$$\frac{Kc}{\mathcal{R}(q,c)} = \frac{1}{M_{w,app}}\left(1 + \frac{q^2 R_G^2}{3}\right) + 2A_2 c \quad (2)$$

In Eq. 2, $K = 4\pi^2 n^2 (dn/dc)^2 / N_A \lambda^4$ is the scattering contrast coefficient ($N_A$ is the Avogadro number) and $A_2$ is the second virial coefficient. The refractive index increments dn/dc of the different solutions were measured on a Chromatix KMX-16 differential refractometer at room temperature. The values of the refractive index increments for the polymer and nanoparticle dispersions are reported in Table I. For the polymers and the nanoparticles in the dilute concentration range (c < 0.2 wt. %), $qR_G \ll 1$ and Eq. 2 reduces to $\mathcal{R}(q,c) = K M_{w,app} c$. This latter equation emphasizes the fact that for small sizes, the Rayleigh ratio does not depend on the wave-vector in the window $6\times10^{-4}$ Å$^{-1}$ - $4\times10^{-3}$ Å$^{-1}$ characteristic for light scattering.

With light scattering operating in dynamical mode, the collective diffusion coefficient D(**c**) was measured in the range c = 0.01 wt. % – 1 wt. %. From the value of D(c) extrapolated at c = 0, the hydrodynamic radius of the colloids was calculated according to the Stokes-Einstein relation, $D_H = k_B T / 3\pi \eta_S D_0$, where $k_B$ is the Boltzmann constant, T the temperature (T = 298 K) and $\eta_0$ (= $0.89\times10^{-3}$ Pa s) the solvent viscosity. The autocorrelation functions of the scattered light were interpreted using both the method of cumulants and the CONTIN fitting procedure provided by the instrument software.

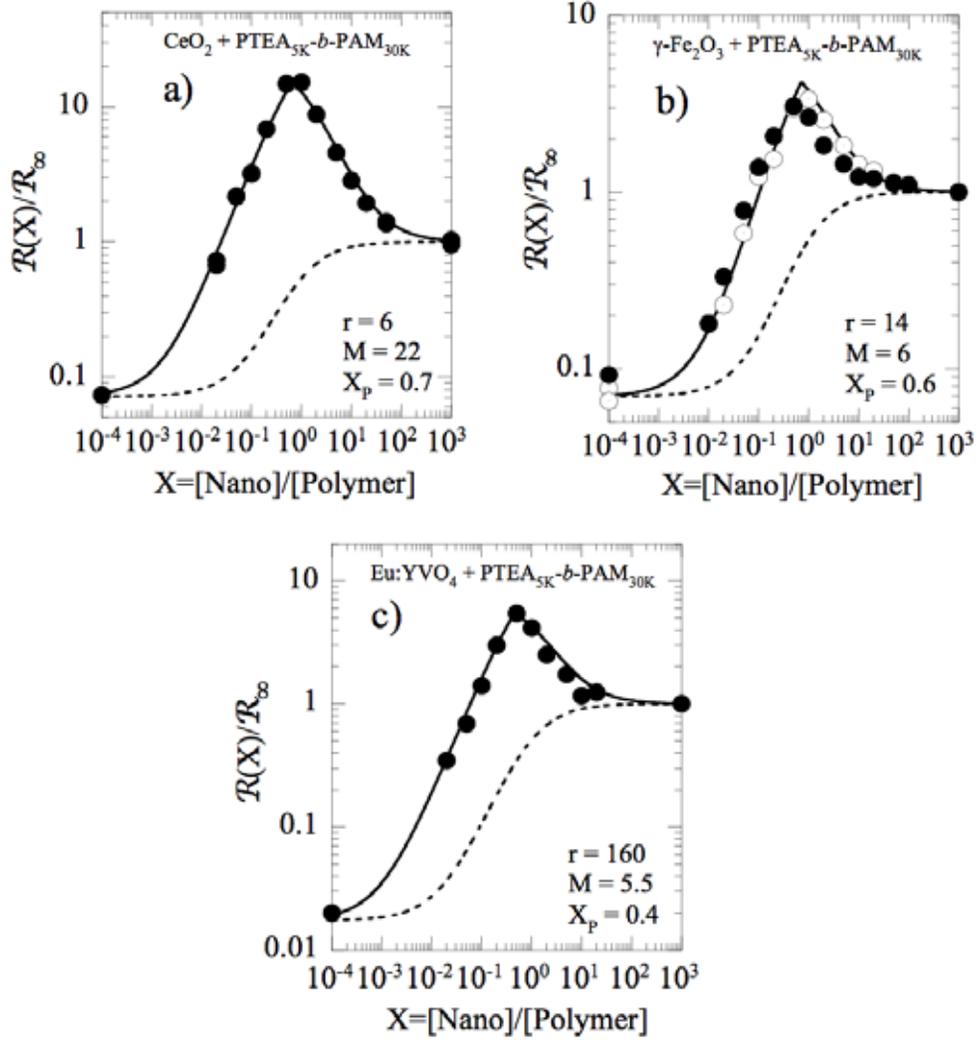

**Figure 1 :** *a)* Normalized Rayleigh ratios $\tilde{R}(X) = R(q_0, c, X)/R_\infty$ obtained at $q_0 = 2\times10^{-3}$ Å$^{-1}$ ($\theta = 90°$) for CeO$_2$ nanoparticles complexed with PTEA$_{5K}$-*b*-PAM$_{30K}$ block copolymers. The total concentration is c = 0.2 wt. % and temperature T = 25° C. *b)* same as in *a)* for γ-Fe$_2$O$_3$ maghemite particles. *c)* Same as in *a)* for Eu:YVO$_4$ particles. The solid lines result from best fit calculations using a stoichiometric association model (Eqs. 7). The dashed lines represent the scattering intensity for unassociated particles and polymers.

## III – Results

Figs. 1a, 1b and 1c displays the normalized Rayleigh ratios $R(q_0,c,X)/R(q_0,c,X=\infty)$ obtained at $q_0 = 2\times10^{-3}$ Å$^{-1}$ ($\theta = 90°$) for the three types of nanoparticles complexed with the PTEA$_{5K}$-$b$-PAM$_{30K}$ block copolymers, with X ranging from $10^{-2}$ to 100 (c = 0.2 wt. %, T = 25° C). There, the pure polymer and the pure nanoparticle solutions have been set at X = $10^{-4}$ and X = 1000 for convenience. For the three systems, the scattered intensity is found to increase steadily with X, to pass through a sharp maximum and then to decrease to 1. The positions of the maxima are $X_P$ = 0.7 ± 0.05, 0.6 ± 0.1 and 0.4 ± 0.05 for the mixed systems prepared with CeO$_2$, γ-Fe$_2$O$_3$ and Eu:YVO$_4$ respectively. As anticipated from the molecular weights of the different species, the Rayleigh ratio of the polymers $R(q_0,c,X=0)$ lies much below that of the nanoparticles, $R(q_0,c,X=\infty) = R_\infty$. The normalization factors in Figs. 1 are respectively, $R_\infty$ = 3.2×10$^{-4}$, 3.2×10$^{-4}$ and 1.3×10$^{-3}$ cm$^{-1}$ for CeO$_2$, γ-Fe$_2$O$_3$ and Eu:YVO$_4$ nanoparticles. Note that the data for γ-Fe$_2$O$_3$/PTEA$_{5K}$-$b$-PAM$_{30K}$ two series of mixed solutions were prepared and that the data agree reasonably well with each other [49,52].

Fig. 2 displays normalized Rayleigh ratios $\tilde{R}(X) = R(q_0,c,X)/R_\infty$ using copolymers with different molecular weights. Fig. 2a describes the behavior of the nanoparticle/polymer system γ-Fe$_2$O$_3$/PTEA$_{11K}$-$b$-PAM$_{30K}$ whereas Fig. 2b that of Eu:YVO$_4$/PTEA$_{2K}$-$b$-PAM$_{60K}$. The results are qualitatively similar to those of Fig. 1. The Rayleigh ratio exhibits a maximum in the intermediate X-range. For the iron oxide with PTEA$_{11K}$-$b$-PAM$_{30K}$, the position of the maximum $X_P$ has shifted to higher X-values ($X_P$ = 1.2) as compared to that of γ-Fe$_2$O$_3$/PTEA$_{5K}$-$b$-PAM$_{30K}$ system (Fig. 1b). On the other hand, $X_P$ (= 0.2) for Eu:YVO$_4$/PTEA$_{2K}$-$b$-PAM$_{60K}$ is now lower than that observed with PTEA$_{5K}$-$b$-PAM$_{30K}$. It is important to note that the complexation between the CeO$_2$ or Eu:YVO$_4$ nanoparticles with the copolymer with the longest polyelectrolyte block, PTEA$_{11K}$-$b$-PAM$_{30K}$ resulted in the formation of micron-size aggregates which settle down rapidly over time. As a consequence, light scattering was not performed on these solutions.

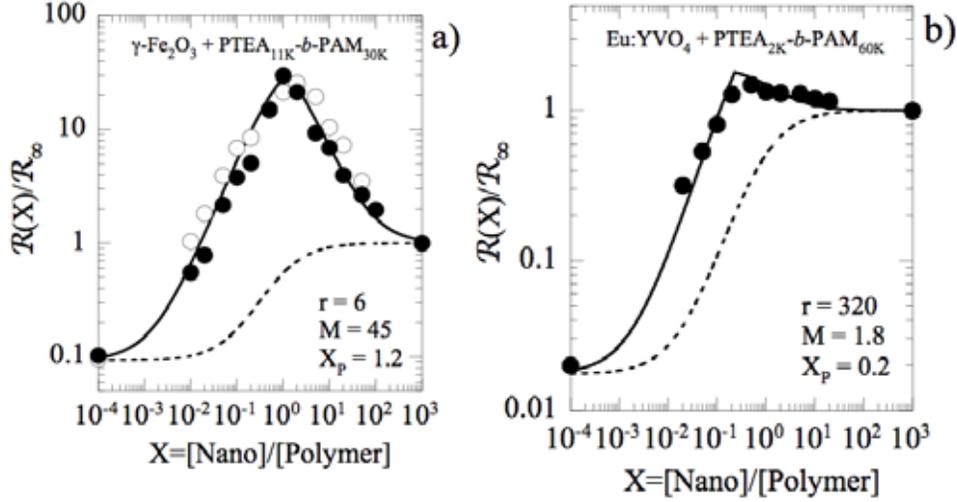

**Figure 2 :** Normalized Rayleigh ratios $\tilde{R}(X) = R(q_0, c, X)/R_\infty$ obtained at $q_0 = 2 \times 10^{-3}$ Å$^{-1}$ ($\theta = 90°$) for $\gamma$-Fe$_2$O$_3$ particles complexed with PTEA$_{11K}$-$b$-PAM$_{30K}$ (*a*) and for Eu:YVO$_4$ particles complexed with PTEA$_{2K}$-$b$-PAM$_{60K}$ (*b*) The solid and dashed lines have the same definitions as in Fig. 1.

Dynamic light scattering performed CeO$_2$/PTEA$_{5K}$-$b$-PAM$_{30K}$ and $\gamma$-Fe$_2$O$_3$/PTEA$_{5K}$-$b$-PAM$_{30K}$ mixed solutions at c = 0.2 wt. % (T = 25 °C) revealed the presence of one or two diffusive relaxation modes, depending on the value of the mixing ratio. Figs. 3a and 3b display the evolution of the hydrodynamic diameters D$_H$ determined for these two modes by the Stokes-Einstein relation. For X > 0.01, the hydrodynamic diameter is much larger than those of the polymers and nanoparticles. It ranges between 60 nm and 80 nm, with an average value at 65 nm for CeO$_2$/PTEA$_{5K}$-$b$-PAM$_{30K}$ and 70 nm for $\gamma$-Fe$_2$O$_3$/PTEA$_{5K}$-$b$-PAM$_{30K}$. For X >> 1, a second mode associated to the single nanoparticles becomes apparent for the two series. In this range, the first order autocorrelation functions were fitted by a double exponential decay. The dashed lines in Figs. 3 mark the coexistence between the two types of colloids. The polydispersity indices obtained from the cumulant analysis were found for these systems in the range 0.10 – 0.25 [53-56]. In the insets of Fig. 3, cryo-TEM images illustrate the core-shell microstructure of the mixed aggregates. The photographs cover spatial fields that are approximately 0.2×0.3 µm$^2$ and display clusters of nanoparticles. For contrast reasons, only the inorganic cores are visible with this technique. The extension of the polymer corona is shown by a circle of diameter D$_H$. The Eu:YVO$_4$/PTEA$_{5K}$-$b$-PAM$_{30K}$ system exhibits a similar behavior (data not shown). At X = 0.05 the hydrodynamic diameter levels off at 120 nm, and at large X-values the results display the coexistence between free particles and aggregates. The findings of a constant hydrodynamic diameter over more than 3 decades in X for the cerium oxide and iron oxide systems, as well as the observation of a coexistence state between free and associated particles at large X suggest the existence of a fixed stoichiometry for the polymer/particle aggregates, regardless of the actual mixing ratio.

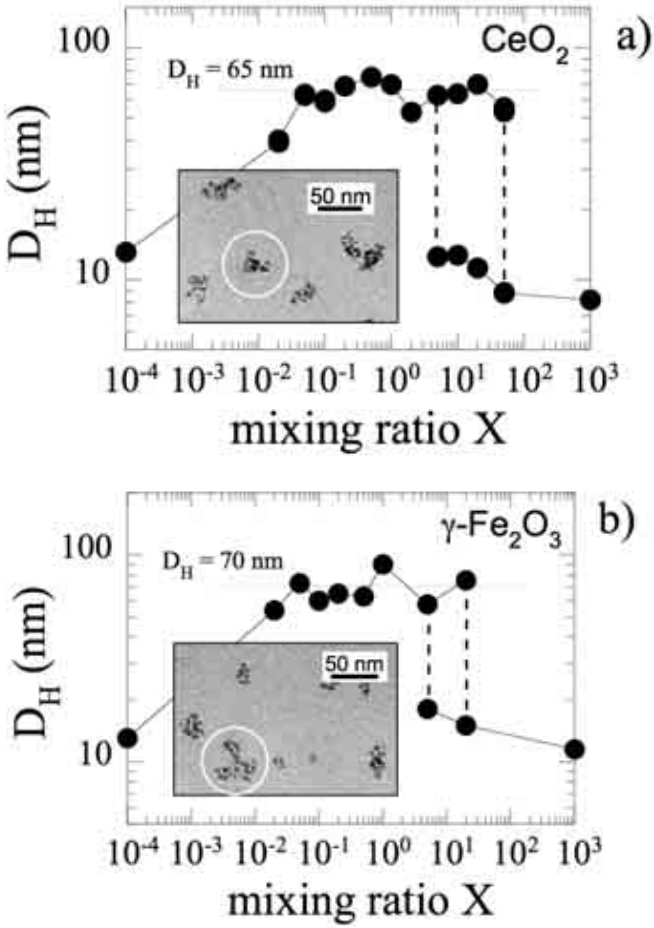

**Figure 3 :** Hydrodynamic diameter $D_H$ as function of the mixing ratio X for mixed dispersions made from PTEA$_{5K}$-$b$-PAM$_{30K}$ block copolymers complexed with cerium oxide nanoparticles (*a*) and with iron oxide nanoparticles (*b*). insets : Cryogenic transmission electron microscopy images of the mixed aggregates. The extension of the polymer corona surrounding the clusters is shown by a circle of diameter $D_H$.

# IV – Modeling the scattered intensity

In this section, we model the X-dependence of the scattering intensity obtained for aqueous dispersions containing oppositely charged polymers and nanoparticles. This approach was inspired by the work of Cabane and coworkers on the adsorption of small colloids (surfactant micelles [57] or cerium oxide nanoparticles [58]) on polymers. More recently, light scattering models were also exploited to describe complex coacervate core micelles obtained by association between oppositely charged polymers [50].

In the experiments described previously, the total concentration of active matter is kept constant. The respective nanoparticle and polymer concentrations in the mixed solutions vary as a function of X as :

$$c^0_{Pol} = \frac{c}{1+X} \; ; \; c^0_{Part} = \frac{cX}{1+X} \qquad (3)$$

The exponent "0" in Eq. 3 refers to the concentrations of all polymers and particles, present in a complexed or in an uncomplexed state. The present model is based of the hypothesis that the solutions contain stoichiometric polymer/particle aggregates at all values of X. The ratio noted r, between polymers and nanoparticles may be expressed as a function of the preferred mixing ratio $X_P$ as :

$$r = \frac{1}{X_P} \frac{M_w^{part}}{M_w^{pol}} \qquad (4)$$

Eq. 4, together with the experimental determination of $X_P$ were utilized in our previous work to obtain an estimate of r [49]. In the present paper, we show that the assumption of a fixed ratio holds over the entire X-range, and that the adjustment of the experimental Rayleigh ratio provides a much better determination of the parameter r. We have not explored here alternative models of association, such as non stoichiometric models [57,58].

Under the assumption of fixed r, $X < X_P$ corresponds then to the domain where the polymers are the major component and where all added nanoparticles participate to the aggregates. For $X > X_P$, it is the reverse : the nanoparticles are in excess and all polymers are used to form the mixed colloids. In the first regime, mixed aggregates are in equilibrium with free polymers, and in the second with free particles. Therefore, a mixed solution prepared at X may comprise all three species. The scattering intensity expresses as the sum of the three contributions :

$$R(q,c,X) = \sum_i K_i c_i(X) \left[ \frac{1}{M_w^i}\left(1 + \frac{q^2 R_{G,i}^2}{3}\right) + 2A_2^i c \right]^{-1} \tag{5}$$

where the index i refers to polymers, mixed aggregates and particles. The definitions of the different parameters in Eq. 5 are identical to those given in the experimental section (Eq. 2). For concentration and wave-vector extrapolated to 0, Eq. 5 simplifies in :

$$R(q \to 0, c, X) = K_{Pol} M_w^{Pol} c_{Pol}(X) + K_{Agg} M_{w,app}^{Agg} c_{Agg}(X) + K_{Part} M_w^{Part} c_{Part}(X) \tag{6}$$

where the equality $\sum_i c_i(X) = c_{Pol}^0 + c_{Part}^0$ insures the mass conservation of the different species. In the following, Eq. 6 will be assumed to describe the intensity measured in the different systems at c = 0.2 wt.% and θ = 90°. For $CeO_2$/PTEA$_{5K}$-$b$-PAM$_{30K}$ and for γ-$Fe_2O_3$/PTEA$_{5K}$-$b$-PAM$_{30K}$, the aggregates are characterized by a radius of gyration $R_G$ ~ 30 nm [59], and the term $\frac{1}{3}q^2 R_G^2$ in Eq. 5 represents only 15 % of the total scattered intensity. For larger aggregates such as those observed with γ-$Fe_2O_3$/PTEA$_{11K}$-$b$-PAM$_{30K}$ (Fig. 2a [49]), the q-dependence of the intensity has to be taken into account.

At low X, the Rayleigh ratio normalized to its value at X = ∞ ($R_\infty = K_{Part} M_w^{Part} c$) expresses as :

$$\tilde{R}(X < X_P) = \tilde{K}_{Pol} m \frac{X_P - X}{X_P(1+X)} + \tilde{K}_{Agg} M \frac{X(1+X_P)}{X_P(1+X)} \tag{7a}$$

The first contribution arises from the unassociated polymers and the second term from the mixed aggregates. Similarly, at large X, the coexistence occurs between the particles and the aggregates and the intensity goes as :

$$\tilde{R}(X > X_P) = \tilde{K}_{Agg} M \frac{1+X_P}{1+X} + \frac{X - X_P}{1+X} \tag{7b}$$

where the notations

$$\tilde{K}_{Pol} = \frac{K_{Pol}}{K_{Part}}, \quad \tilde{K}_{Agg} = \frac{K_{Agg}}{K_{Part}}, \quad m = \frac{M_w^{Pol}}{M_w^{Part}} \text{ and } M = \frac{M_{w,app}^{Agg}}{M_w^{Part}} \tag{8}$$

have been adopted. With these notations, Eq. 4 may be rewritten $r = (mX_P)^{-1}$. It is interesting to note that the normalized scattered intensity in Eqs. 7 does not depend on the total concentration c. The expression should hence be valid at all c in the dilute regime. Eqs. 7 were used to fit the scattering data of Figs. 1 and 2, keeping r (or equivalently $X_P$, Eq. 4) and M (Eq. 8) as adjustable parameters. All others quantities, such as the coupling constants $\tilde{K}_i$ and molecular weights of single constituents were known (Table I). For the mixed aggregates, the values for dn/dc were estimated from the weighted sum of the increments of each component [50,53,54]. The results of the fitting are shown in Figs. 1 and 2 as solid curves and in Table II. The agreement between the model and the data is excellent. There, the position, amplitude and width of the scattering peaks are well accounted for by the predictions of Eqs. 7. In each figures is also displayed for comparison the scattering intensity corresponding to the state where particles and polymers remain unaggregated (dashed lines):

$$\tilde{R}_{\text{UnAgg}}(X) = \frac{\tilde{K}_{Pol}m + X}{1 + X} \qquad (9)$$

Note that in the 5 sets of data of Figs. 1 and 2, the experimental intensities lie all above the predictions for $\tilde{R}_{\text{UnAgg}}(X)$.

The first comment on the results of Table II concerns the shift of the position of the maximum with the molecular weight of the diblock. When the degree of polymerization of PTEA passes from 19 (PTEA$_{5K}$-*b*-PAM$_{30K}$) to 41 (PTEA$_{11K}$-*b*-PAM$_{30K}$) with the γ-Fe$_2$O$_3$ nanoparticles, r decreases by a factor 2. Similarly, with Eu:YVO$_4$ particles, there is again a factor 2 in r between the curves obtained with PTEA$_{2K}$-*b*-PAM$_{60K}$ and with PTEA$_{5K}$-*b*-PAM$_{30K}$. These results suggest that for a given particle/polymer pair, the stoichiometry of the electrostatic complexes *i.e.* the number of polymer per particles is determined by the charge ratio between the two species. The agreement between the data and the predictions of Eqs. 7 indicates that this charge ratio is reached for all aggregates, independently on the actual mixing ratio X. We have estimated the number of positive charges $Q^+$ (coming from the polymers) involved in the complexation process. This number is the product of r and of the degree of polymerization of the polyelectrolyte block. $Q^+$ is found to be +120e, +260e and +2600e for cerium oxide, iron oxide and for europium-doped ytttrium vanadate, respectively (Table II). The value for γ-Fe$_2$O$_3$ $Q^+$ = 260e is lower than that reported in an earlier report (+450e) [49] because in the present approach the scattering data were fitted on the entire mixing range, whereas for the first determination we used the position of the maximum only. For the citrate coated iron oxide particles, the structural charge density was ascertained at -2e nm$^{-2}$ using absorption spectroscopy and conductivity measurements [45]. To the best of our knowledge, this quantity was not derived for the two other nanocrystals. The structural charges of the 6.3 nm γ-Fe$_2$O$_3$ particles amount then at -250e, yielding for the charge ratio in the complexed state the value $Z_P$ = 250/260 = 0.96. A value very close to 1 indicates that the formation of the complexes is accompanied by an almost exact compensation of the electrostatic charges [46,60,61].

| nano-particle | polymer | $\tilde{K}_{Pol}$ | $\tilde{K}_{Agg}$ | $X_P$ | r | M | $Q^+$ |
|---|---|---|---|---|---|---|---|
| CeO$_2$ | PTEA$_{5K}$-b-PAM$_{30K}$ (Fig. 1a) | 0.3 | 0.7 | 0.7 | 6 | 22 | 120 |
| γ-Fe$_2$O$_3$ | PTEA$_{5K}$-b-PAM$_{30K}$ (Fig. 1b) | 0.7 | 0.7 | 0.6 | 14 | 6 | 260 |
| | PTEA$_{11K}$-b-PAM$_{30K}$ (Fig. 2a) | | | 1.2 | 6 | 45 | |
| Eu:YVO$_4$ | PTEA$_{2K}$-b-PAM$_{30K}$ (Fig. 2b) | 1.3 | 1.0 | 0.2 | 320 | 1.8 | 2600 |
| | PTEA$_{5K}$-b-PAM$_{30K}$ (Fig. 1c) | | | 0.4 | 160 | 5.5 | |

**Table II**
List of the parameters derived from fitting the Rayleigh intensities *versus* X using Eqs. 7. The coupling constants $\tilde{K}_{Pol}$ and $\tilde{K}_{Agg}$ are defined as ratios of the actual coupling constants to that of the nanoparticles (Eq. 8). In the remaining columns, $X_P$ denotes the preferred mixing ratio, r the number of polymers per particles (Eq. 4), M the weight-averaged molecular weight of the mixed aggregates and $Q^+$ the number of positive charges per particle involved in the complexation process.

As far as the apparent molecular weight of the complexes is concerned, $M_{w,app}^{Agg}$ is found to range from 3×10$^6$ to 15×10$^6$ g·mol$^{-1}$ for the three different systems listed in Table II [59]. It is calculated by the product M times the molecular weight of the particles (Tables I and II). As already mentioned, for the systems such as γ-Fe$_2$O$_3$/PTEA$_{11K}$-b-PAM$_{30K}$ where the aggregates are in the range 200 nm, the values for $M_{w,\text{app}}^{Agg}$ are underestimated by the model. From the data in Table II, we observe a strong dependence of the molecular weight of the mixed aggregates with respect to the molecular weight of the polyelectrolyte block. The parameter M increases by a factor around 8 (resp. 3) for γ-$Fe_2O_3$ (resp. Eu:$YVO_4$) systems when the charged block is multiplied by 2. Similar results were obtained for the complexation of cationic surfactant micelles with anionic/neutral block copolymers [48]. Note finally that the molecular weight derived in this way are in qualitative agreement with those obtained from size distribution of the inorganic cluster, as deduced from cryo-TEM experiments.

## V – Concluding Remarks

In the present paper it is shown that the mixing protocol for oppositely charged particles and polymers is appropriate to study the formation of electrostatic complexes. This methodology has allowed us to explore a broad range in mixing ratios and nevertheless to keep the total concentration in the dilute regime. With dispersions remaining in the dilute regime, a quantitative interpretation of the light scattering data has been made possible. The protocols were applied to three types of nanoparticle dispersions, namely $CeO_2$, γ-$Fe_2O_3$ and Eu:$YVO_4$ nanocrystals and three molecular weight of the copolymer poly(trimethylammonium ethylacrylate)-*b*-poly(acrylamide). The use of cationic-neutral copolymers was required to prevent the polymers and nanoparticles to phase separate upon mixing, as this is anticipated from dispersions of oppositely charged species. With light scattering, we have found a unique behavior for the evolution of the scattering intensity as a function of the mixing ratio. This intensity exhibited a sharp and prominent peak, which could be reproduced using a stoichiometric model of association. For 5 pairs of polymers and particles, the agreement between the results and the model was remarkable. These findings have allowed us to conclude that the stoichiometry of the mixed aggregates is controlled by the electrostatic interactions between the opposite charges. We finally suggest that the protocols and the association model developed could be applied to other types of non-covalent and reversible binding. One can think for instance of oppositely charged species below the critical concentration of the phase separation [14,15,17], or of colloids and polymers with weak electrostatic or H-bonding [58] interactions. This methodology could be also extended to non-stoichiometric models for the determination of adsorption isotherms of ligands, proteins or macromolecules.

**Acknowledgements** : We thank Yoann Lalatonne, Nicolas Schonbeck, Delphine El Karrat for their support in the light scattering experiments. We are also indebted to Mikel Morvan, Amit Sehgal (Complex Fluids Laboratory, Bristol PA, USA), Olivier Sandre (Laboratoire des Liquides Ioniques et Interfaces Chargées, Université Pierre et Marie Curie-Paris 6, France) for fruituil discussions and Mathias Destarac (Centre de Recherches d'Aubervilliers, France) for providing us with the polymers. This research is supported by Rhodia.